\documentclass[preprint]{appolb}
\usepackage{graphicx}
\usepackage[utf8]{inputenc}
\usepackage{hyperref}
\hypersetup{colorlinks=false, linkcolor=black, filecolor=black, urlcolor=black}
\usepackage{breakurl} % \burl{url}

\begin{document}

\title{\texttt{MBsums} -- a Mathematica package for the representation of Mellin-Barnes integrals by multiple sums%
\thanks{Corresponding author: Tord Riemann, E-mail: tordriemann@gmail.com. Part of a talk presented by J. Gluza at the XXXIX International 
Conference of Theoretical Physics
“Matter to the Deepest”, Ustron, Poland, September 13–18, 2015. Extended version of the contribution to the proceedings in Acta 
Physica Polonica 46 (2015).
}}
% % \author{
% % Micha\l{} Ochman\address{Deutzer Technische Kohle GmbH,
% % Goethestraße 35,
% % %Lindenallee 16,
% %  15738 Zeuthen, Germany 
% % %Eichwalde, Germany
% % }
% % \and
% % Tord Riemann\address{K\"onigs Wusterhausen, Germany}}
\author{Micha\l{} Ochman and 
Tord Riemann\address{Deutsches Elektronen-Synchrotron, DESY, Platanenallee 6, 15738 Zeuthen, Germany}}

%\date{}

\maketitle

\begin{abstract}
Feynman integrals may be represented by the Mathematica packages \verb|AMBRE| and \verb|MB| as multiple Mellin-Barnes integrals.
With the Mathematica package \verb|MBsums| we transform these Mellin-Barnes integrals into multiple sums.
\end{abstract}

\section{Introduction}\sloppy
In recent years, there was remarkable progress in the development of (semi-)automatized software for the numerical calculation of 
arbitrary, complicated Feynman integrals. 
Basically, two approaches are advocated. One relies on sector decomposition. For an introduction given at this conference and for further 
references see \cite{borowka-ustron-break:2015}. 
We will report on the other approach, based on Mellin-Barnes representations 
\cite{gluza-ustron-break:2015,zbMATH02640947,Usyukina:1975yg,Boos:1990rg,Smirnov:1999gc,
Tausk:1999vh,mbtools,Czakon:2005rk,Smirnov:2009up,Gluza:2007rt,Gluza:2010rn}. 
When \cite{Czakon:2005rk} appeared in 2005, several unsolved problems of different complexity existed. We mention non-planar diagrams, the 
massive case, multi-loop tensor integrals, Minkowskian kinematics. For all of the items, progress is reported in 
\cite{gluza-ustron-break:2015}, based on the source-open software AMBRE/MB \cite{Dubovyk:2015yba,Katowice-CAS:2015,mbtools}.
An alternative is the direct analytical evaluation of MB-integrals. 
This is difficult.
But in view of recent progress in algebraically summing up infinite sums by the LINZ group's
Computer algebra algorithms for nested sums and products, one might hope to achieve a breakthrough \cite{linz-group:2015}; 
certainly only if the result leads to appropriate classes of functions. 
The idea is to apply the Linz group's algorithms (as well as those of others, e.g. \cite{Moch:2005uc}) to sums of residues after 
applying Cauchy's theorem \cite{Riemann:2012linzAPP}.   
A first attempt was reported in \cite{Blumlein:2014maa}.

The automatized derivation of the multiple sums for a given MB-integral is certainly the easier part of the task, but it is the first step. 
We report here on a first version of the  Mathematica program  MBsums \cite{ochman-mbsums:2015} for transforming MB-representations for 
Feynman integrals into multiple sums. The licence conditions of the source-open package are those formlulated in the CPC non-profit use 
licence agreement of the Computer Physics Communications Program 
Library \cite{cpc-licence:7Oct2015}. 
% Collegues who do not want to respect this licence, should not use the package. 
The authors expect that the potential users read and follow the licence
agreement when using this code.
\section{The Mathematica package {\tt MBsums}}

The package \verb|MBsums| transforms 
Mellin-Barnes integrals 
% (MB-integrals, \cite{zbMATH02640947,Usyukina:1975yg,Boos:1990rg,Smirnov:1999gc,
% Tausk:1999vh,mbtools,Czakon:2005rk,Smirnov:2009up,Gluza:2007rt,%
% Gluza:2010rn}) ({\bf Cite also LL2014 and \verb|MBsums v0.9|?}) 
into sums, by closing the integration contours and calculating the integrals by the residue 
theorem, i.e. by constructing sums over all residues inside the contours.
The current version of \verb|MBsums| is 1.0. The package \verb|MBsums| works with Wolfram Mathematica 7.0 and later.

In order to obtain a sum from an MB-integral the user should use the \verb|MBIntToSum| function of \verb|MBsums|:
\begin{equation} \label{exampleMBgen} 
\verb|MBIntToSum[int,{},contours]|  
\end{equation}
or
\begin{equation} \label{exampleMBgen2} 
\verb|MBIntToSum[int,kinematics,contours]|,
\end{equation}
where \verb|int| is the MB-integral in the form as it is denoted in the Mathematica  package \verb|MB| \cite{Czakon:2005rk}%
\footnote{%
The package {\tt MBsums} uses notations of the package {\tt MB}, but can be run also independently.}:
\begin{equation} \label{int} 
 \verb|int| = \verb|MBInt[f,{{eps -> 0},{z1 -> c1, z2 -> c2, ... , zD -> cD}]|
\end{equation}
which corresponds to
\begin{equation} \label{int2} 
 \verb|int| = \frac{1}{(2\pi i)^d}\int_{-i\infty+c_1}^{i\infty+c_1}
\cdots \int_{-i\infty+c_D}^{i\infty+c_D} f\prod_{k=1}^D\mathrm{d}z_k.
\end{equation}
The integrand f can have the form
\begin{equation}
f=\sum_j f_j
\end{equation}
and each $f_j$ is assumed to be of the form of
\begin{equation}
f_j = \xi_j \frac{\prod_m\Gamma(N_m^{(j,1)}) \prod_m\Psi^{(n)}(N_m^{(j,2)})}
{\prod_m\Gamma(N_m^{(j,3)}) } \prod_m r_m^{N_m^{(j,4)}},
\end{equation}
where $\Psi^{(0)}(z)=\mathrm{d}\ln(\Gamma(z))/\mathrm{d}z$, $\Psi^{(n>0)}(z)=\mathrm{d}^{n}\Psi^{(0)}(z)/\mathrm{d}z^{n}$, $r_i$ are free parameters (usually kinematic parameters) in \verb|int| and $\xi_j$ is a factor independent of $z$-variables. The
\begin{equation}  
N_m^{(j,k)}=\sum_i \alpha_{m,i}^{(j,k)} z_i+\beta_m^{(j,k)}+\gamma_m^{(j,k)}\epsilon,
\end{equation}
where $\alpha_{m,i}^{(j,k)}$, $\beta_m^{(j,k)}$, $\gamma_m^{(j,k)}$ are rational numbers and $\epsilon$ (\verb|eps|) is an infinitesimal 
dimensional shift,  e.g. arising from $d = 4 - 2\epsilon$. 
All the singularities of the integrand of the MB-integral \verb|f| are due (and only due) to \verb|Gamma| and \verb|PolyGamma| functions.

The values of \verb|c1, c2, ... , cD| are converted to rational numbers by \verb|MBIntToSum| before calculations.

Let us now focus on the case when the list  \verb|kinematics| is empty, i.e. we will consider (\ref{exampleMBgen2}). The list  
\verb|contours| has the form
\begin{equation}  
 \verb|contours|  = \verb|{z1 -> L/R, z2 -> L/R, ... , zD -> L/R}|.
\end{equation}
The order of the $z$-variables defines the order of integrations chosen by the user (from left to right). The {\tt L} ({\tt{R}}) means that 
the 
contour will be closed to the left (right). The {\tt L}/{\tt{R}} choice made by the user can be changed if \verb|kinematics| is not empty 
and this will be covered later. The output of \verb|MBIntToSum| in (\ref{exampleMBgen2}) is of the form 
\begin{equation} \label{Sum}
\verb|{MBsum_1, MBsum_2, ... , MBsum_Q}|
\end{equation}
where 
\begin{equation}
\verb|MBsum_i| = \verb|MBsum[Sum_Coefficient_i,Conditions_i,List_i]|
\end{equation}
 represents a sum with summand \verb|Sum_Coefficient_i| that has non-negative indices given in the list \verb|List_i|, and 
\verb|Conditions_i| are conditions on those indices. The complete answer is the sum of all \verb|MBsum_i| in the list.

The list  \verb|kinematics| has the form
\begin{equation}  
 \verb|kinematics|  = \verb|{r_1 -> v_1, r_2 -> v_2, ... , r_K -> v_K}|,
\end{equation}
where \verb|r_i| are free parameters (usually kinematic parameters) in \verb|int| and \verb|v_i| are values of \verb|r_i|. If 
\verb|kinematics| is not empty then \verb|MBIntToSum| will try to change the {\tt L}/{\tt{R}} choice made by the user in the list  
\verb|contours| in order to obtain sums that have good asymptotic behaviour at given values of \verb|r_i|. In any case the user is informed 
how the contour was closed. This will be explained later in detail. The values of \verb|v_i| are converted to rational numbers by 
\verb|MBIntToSum| before calculations. The user can turn off all messages printed by \verb|MBIntToSum| by typing \verb|MBsumsInfo=False| and 
turn them on by typing \verb|MBsumsInfo=True|.

In addition we provide function \verb|DoAllMBSums[sums,nmax,kinematics]| that sums the \verb|sums| in the form (\ref{Sum}). The \verb|nmax| 
is the maximal value of each index, the minimal value is given by conditions on indices. The list  \verb|kinematics| is as above and may be 
empty. We used Wolfram Mathematica function \verb|ParallelMap| inside \verb|DoAllMBSums| to sum individual sums in the list \verb|sums| in 
parallel.

\section{Obtaining the sums}

Let us now shortly explain how we obtain the sums. We point out the most important ingredients in our algorithm. Let us focus on the case 
when the list  \verb|kinematics| is empty, i.e. we will consider (\ref{exampleMBgen2}). The MB-integral is in the form as it was denoted in 
(\ref{int}). Let us now assume  that the user has chosen as first integration the \verb|z2->L|. As a first step we form a list, which we 
call  NegArgsDoC, of arguments of the \verb|Gamma| and \verb|PolyGamma| functions in the numerators that give residues for $Re(z2)<c2$ 
%We do it in the following way. For each argument $G$ of the \verb|Gamma| and \verb|PolyGamma| functions in the numerators we do
%\begin{enumerate}
(see (\ref{int}), the remaining contours are seen as a straight lines). We call that list NegArgsDoC. 
Next w consider all possible cases: When all   
\verb|Gamma| and \verb|PolyGamma| functions that have arguments in NegArgsDoC contribute to a residue at the same time, and when only some 
subset of them contributes to a residue at the same time. We consider all possible subsets of NegArgsDoC. Additionally we have to be 
careful when some \verb|Gamma| %and \verb|PolyGamma| 
functions in the denominator become singular at some points. If we have terms like \verb|Gamma[2 z2]|, then the poles are at $z2=-n/2$ and 
we consider there 2 cases: $n=2n'$ and $n=2n'+1$, where $n$ and $n'$ are non-negative integers. Similarly we proceed with arbitrary $M 
\times z2$ terms, where $M$ is some integer value and in general with all $n/M$ terms which appear together with integration variable in 
the arguments of the \verb|Gamma| and \verb|PolyGamma| functions. So we produce a list of cases
\begin{equation}  \label{a}
\{
\{f_1^{(1)},c_1^{(1)}\},
\{f_2^{(1)},c_2^{(1)}\},...,\{f_{K_1}^{(1)},c_{K_1}^{(1)}\}\},
\end{equation}
where $f_i^{(1)}$  are expressions after taking residues of f and $c_i^{(1)}$ are conditions on the index that numerates terms (residues) 
in $f_i^{(1)}$.  We obtain a list of $K_1$ elements after integrating over z2. Let us now assume  that the user has chosen as second 
integration variable \verb|z5->R|. Then we repeat the whole procedure on each $f_i^{(1)}$ taking into account conditions $c_i^{(1)}$. 
So we produce analogous to (\ref{a}) a list of cases %of K_2 elements, 
\begin{equation}  
\{
\{f_1^{(2)},c_1^{(2)}\},
\{f_2^{(2)},c_2^{(2)}\},...,\{f_{K_2}^{(2)},c_{K_2}^{(2)}\}\}.
\end{equation}
We repeat the whole procedure for each integration variable.

\section{Contours and convergent sums}
Let us now shortly explain how we obtain the sums if the list \verb|kinematics| is not empty. We follow the order of integration given in 
the list  \verb|contours|. Our aim is to determine the {\tt L}/{\tt{R}}  such that we obtain sums that have good asymptotic behaviour at 
given values of \verb|r_i| in the list \verb|kinematics|. We do it in the following way. At each integration step $s$ we analyse the 
expressions $f_i^{(s)}$ that are to be integrated over some $z_C$. Each $f=f_i^{(s)}$ we decompose as
\begin{equation}
f=\sum_j g_j
\end{equation}
and each $g=g_j$ is of the form 
\begin{equation}
g=r_1^{a_1 }r_2^{a_2 }\cdots r_K^{a_K }F\,,\qquad
a_j={\sum_i a_{j,i} z_i},
\end{equation}
where  $r_i$ are the kinematic parameters in the list \verb|kinematics| and $F$ contains the rest of $g$. If we integrate over $z_C$, we 
consider
\begin{equation}
c^{z_C}\,,\qquad c=r_1^{a_{1,C} }r_2^{a_{2,C} }\cdots r_K^{a_{K,C} } .
\end{equation}
The value of $c$ is calculated.  \verb|MBIntToSum| prints the error message:
\begin{eqnarray}
& & \verb|Found c = |c\verb| (not a number): please complete kinematic's list.|\nonumber
\end{eqnarray}
for each $g_i$ in each $f_i^{(s)}$ when $c$ is symbolic (not a number) and at the end \verb|MBIntToSum| prints 
\begin{eqnarray}
& & \verb|Unable to find correct contour for |z_C\verb|.|\nonumber
\end{eqnarray}
and returns \{\}. 
The user should complete the list \verb|kinematics|.

For each $g_i$ in each $f_i^{(s)}$ it is returned {\tt L} if $|c|>1$ or {\tt R} if $|c|<1$ indicating how to close the contour or \{\} if 
$|c|=1$.

If for each $g_i$ in each $f_i^{(s)}$ it is returneded {\tt L} ({\tt R}) or \{\} then the contour for $z_C$ will be closed to left (right) 
if it is returned at least one {\tt L} ({\tt R}).

If for each $g_i$ in each $f=f_i^{(s)}$ it is returned \{\} then the choice of user given in the list  \verb|contours| is taken.

 If for some $g_i$ it is returned {\tt R} and for some $g_j$ it is returned {\tt L} then  \verb|MBIntToSum| prints the error message:
\begin{eqnarray}
& & \verb|Unable to find correct contour for |z_C\verb|.|\nonumber
\end{eqnarray}
and returns \{\}.  Otherwise we compute the sums as described above. We repeat the whole procedure for each integration variable. We stress that this procedure as described above does not always give convergent sums.

There are MB-integrals for which no convergent sums can be found. One such example is the following MB-integral:
\begin{equation}
B_1 = \frac{1}{2\pi i}\int_{-i\infty}^{i\infty} \mathrm{d}z \Gamma^2(1 + z)\Gamma^2(1 - z)\,.
\end{equation}
Here we can apply the first Barnes lemma \cite{zbMATH02640947} and obtain $B_1 = 1/6$, but the reader can check that indeed the infinite series of residues diverge both for $\mathrm{Re}\, z>0$ and $\mathrm{Re}\, z<0$.

Consider the following integral (see also \cite{zbMATH02640947}):
\begin{equation}
B_x = \frac{1}{2\pi i}\int_{-i\infty}^{i\infty} \mathrm{d}z x^z \Gamma^2(1 + z)\Gamma^2(1 - z),\ x>0\land x\neq 1.
\end{equation}
Closing the contour to the right ($\mathrm{Re}\, z>0$) gives the following series
\begin{equation}
s_R = -\sum_{n=1}^{\infty}n x^{n} (2 + n \ln(x)) ,
\end{equation}
convergent for $0<x<1$, while closing the contour to the left ($\mathrm{Re}\, z<0$) gives the following series
\begin{equation}
s_L = -\sum_{n=1}^{\infty}n x^{-n} (2 - n \ln(x)) ,
\end{equation}
convergent for $x>1$. Both $s_L$ and $s_R$ give the same formula after summing up, that is
\begin{equation}
s_{LR} = \frac{x (2 - 2 x + (1 + x) \ln(x))}{(x-1)^3},\ x>0\land x\neq 1 ,
\end{equation}
so $B_x=s_{LR}$ and $\lim_{x\to1}B_x=B_1$.

\section{Simplification of sums}

We also provide function \verb|SimplifyMBsums[sums]| that takes each \verb|MBsum[]| in the list \verb|sums| in the form (\ref{Sum}) and 
simplifies it. The output of \verb|SimplifyMBsums| is of the form (\ref{Sum}). If \verb|MBsum[f,c,l]| is the sum to be simplified then 
\verb|SimplifyMBsums| uses the Mathematica function \verb|Reduce| to simplify the conditions \verb|c|. If \verb|Reduce| returns 
-- after applying Mathematica function \verb|LogicalExpand| to \verb|c| --
for 
\verb|c| an
answer in the form
% (after applying Mathematica function \verb|LogicalExpand| to \verb|c|) 
\verb-c_1||c_2||...||c_R-, then we decompose 
\verb|MBsum[f,c,l]| into 
$$\verb|{MBsum[f,c_1,l],MBsum[f,c_2,l],...,MBsum[f,c_R,l]}|$$.
If some \verb|MBsum[f,c_i,l]| contains conditions \verb|c_i| like
\begin{verbatim}  
n1 == C[1] && n2 == C[1] + C[2]
\end{verbatim}
where \verb|n1|, \verb|n2| are summation indices in \verb|MBsum[f,c_i,l]|, then we make the substitutions \verb|n1 -> C[1]| and 
\verb|n2 -> C[1] + C[2]| in \verb|f| and update \verb|c_i| and \verb|l|  in 
\verb|MBsum[f,c_i,l]|; \verb|C[1]| and \verb|C[2]| are new summation indices.
We assume that \verb|C[]| are integers. Next, we unify the names of  indices. Finally, we combine sums having the 
same conditions. It can happen e.g. that we obtain 4-dimensional sums from a 3-dimensional sum after applying \verb|SimplifyMBsums|. 
To 
illustrate this, let us consider the following example. Let \verb|c| be
\begin{verbatim}  
c = n3 >= 0 && n2 >= 1 && n1 >= 0 && 
    n2 >= 1 + n3 && 1 + n1 <= n2
\end{verbatim}
then
\begin{verbatim}  
Reduce[c, {n1, n2, n3}, Integers]
\end{verbatim}
gives
\begin{verbatim}  
Element[C[1] | C[2] | C[3] | C[4], Integers] && 
C[1] >= 0 && C[2] >= 0 && C[3] >= 0 && C[4] >= 0 && 
n1 == C[1] + C[2] && 
n2 == 1 + C[1] + C[2] + C[3] + C[4] && 
n3 == C[2] + C[3]
\end{verbatim}

In addition we provide function \verb|MergeMBsums[sums]| that only combines sums in the list \verb|sums| (in the form (\ref{Sum})) having 
the same conditions.

\section{Examples}

In all the examples we will proceed as follows. We start with Feynman integral $I$ in $d = 4 - 2\epsilon$ dimensions. First, 
the MB-representation for $I$ is obtained using the Mathematica  package \verb|AMBRE| v.1.2 
\cite{Gluza:2007rt,Gluza:2010rn}.
Next the analytical continuation $\epsilon\to 0$ as well as the expansion of the 
integrals around $\epsilon=0$ is obtained using the Mathematica  package \verb|MB| v.1.2 \cite{Czakon:2005rk,mbtools}.
In some examples, we also used the Mathematica  package \verb|barnesroutines| v.1.1.1 by  David A. Kosower \cite{mbtools} in order to 
reduce the dimensionality of the MB-integrals. 

\subsection{Massive QED on-shell one-loop box diagram}

The first example is the massive QED one-loop box diagram with two photons in the $t$-channel; see the example file \verb|1lbox.nb| 
~\cite{gluza-ustron-break:2015}.
The input integral is, with kinematics $p_i^2=m^2$ for $i=1,2,3,4$ and $p_1+p_2=-p_3-p_4$:
\begin{equation}
I=  e^{\gamma_E \epsilon}\int \frac{d^{d}k_{1}}{i\pi^{d/2}}  
  \frac{1}{(k_{1}^2-m^2) (k_{1}+p_{1})^2 (k_{1}-p_{3})^2 [(k_{1}+p_{1}+p_{2})^2-m^2]}.
\end{equation}
Using \verb|AMBRE| one obtains the following MB-representation for finite $\epsilon$:
\begin{eqnarray} \label{box}
I & = & e^{\gamma_E \epsilon}\frac{(-t)^{-2 - \epsilon}}{(2\pi i)^2 \Gamma(-2 \epsilon)}\int_{-i\infty}^{i\infty} \mathrm{d}z_1\int_{-i\infty}^{i\infty} \mathrm{d}z_2
  ( (-x)^{z_1} y^{z_2} \frac{\Gamma^2(1 + z_2) \Gamma(-z_2)}{\Gamma(2 + 2 z_2)} 
   \nonumber\\ &   & \times \Gamma(-z_1)\Gamma^2(-1 - \epsilon - z_1 - z_2) \Gamma(2 + \epsilon + z_1 + z_2) \Gamma(
    2 + 2 z_1 + 2 z_2)) \nonumber\\
\end{eqnarray}
with $x = m^2/t$, $y = s/t$,  $(p_1+p_2)^2=s$, $(p_1+p_3)^2=t$. %
%\footnote
 {The MB-representation (\ref{box}) is the same, up to irrelevant 
factors, as that %on page 30 
in \cite{Riemann:2012linz}.}
See also the sums there, compare also with the MB-representation in \verb|example3.nb| 
\cite{Katowice-CAS:2015}, described in \cite{Gluza:2007rt}.
Next one performs the analytical continuation of (\ref{box}) for $\epsilon\to 0$ 
and the expansion of the resulting integrals around $\epsilon=0$ up to $\epsilon^1$ using \verb|MB|:
\begin{verbatim}
{MBint[((-x)^z1*y^(-1 - z1)*Gamma[-z1]^3*Gamma[1 + z1]*
    (-6 + 3*eps^2*EulerGamma^2 + 4*eps^2*Pi^2 + 
     6*eps*Log[-t] - 3*eps^2*Log[-t]^2 - 
     6*eps^2*EulerGamma*Log[y] + 3*eps^2*Log[y]^2 + 
     12*eps^2*PolyGamma[0, -2*z1]^2 + 
     12*eps^2*PolyGamma[0, -z1]^2 + 6*eps^2*EulerGamma*
      PolyGamma[0, 1 + z1] - 6*eps^2*Log[y]*
      PolyGamma[0, 1 + z1] + 3*eps^2*PolyGamma[0, 1 + z1]^
       2 - 12*eps^2*PolyGamma[0, -z1]*(EulerGamma - 
       Log[y] + PolyGamma[0, 1 + z1]) + 
     12*eps^2*PolyGamma[0, -2*z1]*(EulerGamma - Log[y] - 
       2*PolyGamma[0, -z1] + PolyGamma[0, 1 + z1]) - 
     12*eps^2*PolyGamma[1, -2*z1] + 
     6*eps^2*PolyGamma[1, -z1] + 
     3*eps^2*PolyGamma[1, 1 + z1]))/
   (6*eps*t^2*Gamma[-2*z1]), {{eps -> 0}, {z1 -> -1/2}}], 
 MBint[(-2*eps*(-x)^z1*y^z2*Gamma[-z1]*
    Gamma[-1 - z1 - z2]^2*Gamma[-z2]*Gamma[1 + z2]^2*
    Gamma[2*(1 + z1 + z2)]*Gamma[2 + z1 + z2])/
   (t^2*Gamma[2 + 2*z2]), {{eps -> 0}, 
   {z1 -> -1/2, z2 -> -1/4}}]}
\end{verbatim}
The expression agrees numerically with the MB-representations in Eq.~(4.25) of \cite{Smirnov:2004} and in Eq.~(4.67) 
of \cite{Fleischer:2006ht} after analytical continuation and expansion in $\epsilon$ by \verb|MB|.

We would now like to reproduce the representation by sums of the MB-integrals above. We aim at sums to be convergent at 
\begin{equation} \label{Lk}
\verb|Lk = {x -> -1/10, y -> 1/50, t -> -10}|.
\end{equation}
We start with the first, one-dimensional  integral on the list, which we denote by \verb|dim1int|. Then 
$$\verb|dim1sum = MBIntToSum[dim1int, Lk, {z1 -> L}]| $$ gives
\begin{verbatim}
{MBsum[-(y^n1*n1!^2*(-4 + 3*eps^2*Pi^2 + 
      2*eps^2*HarmonicNumber[n1]^2 - 
      2*eps^2*HarmonicNumber[n1, 2] + 4*eps*Log[-t] - 
      2*eps^2*Log[-t]^2 - 4*eps^2*HarmonicNumber[n1]*
       Log[-x] + 2*eps^2*Log[-x]^2))/(4*(-1)^n1*eps*t^2*
    (-x)^n1*x*(1 + 2*n1)!), n1 >= 0, {n1}]}
\end{verbatim}
together with the message how the contour was closed: $$\verb|z1->L ( Re z1 < -1/2 )|$$. Here, the original choice how to close the 
contour is unchanged. The user can check that the same result can be obtained with \verb|dim1sum = MBIntToSum[dim1int, {}, {z1 -> L}]|.
 One can now check the result numerically: $$\verb|DoAllMBSums[dim1sum, 50, Lk] // N|$$ gives
\begin{verbatim}
0.222878 - 0.0967945/eps + 0.709316 eps
\end{verbatim}
in agreement with
\begin{verbatim}
{0.222878 - 0.0967945/eps + 0.709316 eps, 0.}
\end{verbatim}
from \verb|MBintegrate[{dim1int}, Lk]//N|.

Next, we do the  two-dimensional integral on the list, which we denote by \verb|dim2int|. Then
\begin{verbatim}
dim2sum = MBIntToSum[dim2int, Lk, {z1 -> L, z2 -> L}]
\end{verbatim}
gives
\begin{verbatim}
{MBsum[-(eps*(-x)^(n1 - n2)*y^n2*(-1 + 2*n1)!*n2!*
     (Pi^2 + 6*HarmonicNumber[n1]*HarmonicNumber[
        -1 + n1 - n2] - 12*HarmonicNumber[-1 + 2*n1]*
       HarmonicNumber[-1 + n1 - n2] + 
      3*HarmonicNumber[-1 + n1 - n2]^2 + 
      6*HarmonicNumber[n1]*HarmonicNumber[n2] - 
      12*HarmonicNumber[-1 + 2*n1]*HarmonicNumber[n2] - 
      3*HarmonicNumber[n2]^2 - 12*HarmonicNumber[n1]*
       HarmonicNumber[1 + 2*n2] + 
      24*HarmonicNumber[-1 + 2*n1]*HarmonicNumber[
        1 + 2*n2] + 12*HarmonicNumber[n2]*
       HarmonicNumber[1 + 2*n2] - 
      12*HarmonicNumber[1 + 2*n2]^2 + 
      3*HarmonicNumber[-1 + n1 - n2, 2] + 
      3*HarmonicNumber[n2, 2] - 12*HarmonicNumber[1 + 2*n2, 
        2] - 6*HarmonicNumber[n1]*Log[-x] + 
      12*HarmonicNumber[-1 + 2*n1]*Log[-x] - 
      6*HarmonicNumber[-1 + n1 - n2]*Log[-x] + 
      3*Log[-x]^2 + 6*HarmonicNumber[n1]*Log[y] - 
      12*HarmonicNumber[-1 + 2*n1]*Log[y] - 
      6*HarmonicNumber[n2]*Log[y] + 
      12*HarmonicNumber[1 + 2*n2]*Log[y] - 3*Log[y]^2))/
   (3*(-1)^(3*n1)*t^2*x*n1!*(-1 + n1 - n2)!*(1 + 2*n2)!), 
  2*n1 >= 1 && n2 >= 0 && n1 >= 1 + n2, {n1, n2}], 
 MBsum[(2*(-1)^(-2*n1 - n2)*eps*(-x)^(n1 - n2)*y^n2*
    (-1 + 2*n1)!*n2!*(-n1 + n2)!*(HarmonicNumber[n1] - 
     2*HarmonicNumber[-1 + 2*n1] + HarmonicNumber[
      -n1 + n2] - Log[-x]))/(t^2*x*n1!*(1 + 2*n2)!), 
  2*n1 >= 1 && n2 >= 0 && n1 <= n2, {n1, n2}], 
 MBsum[(eps*(-x)^n1*y^(-1 - n1 + n2)*(-1 - n1 + n2)!*
    (-1 + 2*n2)!*(Pi^2 + HarmonicNumber[n2]^2 - 
     2*HarmonicNumber[n2]*HarmonicNumber[-1 - n1 + n2] + 
     HarmonicNumber[-1 - n1 + n2]^2 - 4*HarmonicNumber[n2]*
      HarmonicNumber[-1 + 2*n2] + 
     4*HarmonicNumber[-1 - n1 + n2]*HarmonicNumber[
       -1 + 2*n2] + 4*HarmonicNumber[-1 + 2*n2]^2 + 
     4*HarmonicNumber[n2]*HarmonicNumber[
       -1 - 2*n1 + 2*n2] - 4*HarmonicNumber[-1 - n1 + n2]*
      HarmonicNumber[-1 - 2*n1 + 2*n2] - 
     8*HarmonicNumber[-1 + 2*n2]*HarmonicNumber[
       -1 - 2*n1 + 2*n2] + 
     4*HarmonicNumber[-1 - 2*n1 + 2*n2]^2 + 
     HarmonicNumber[n2, 2] - HarmonicNumber[-1 - n1 + n2, 
      2] - 4*HarmonicNumber[-1 + 2*n2, 2] + 
     4*HarmonicNumber[-1 - 2*n1 + 2*n2, 2] - 
     2*HarmonicNumber[n2]*Log[y] + 
     2*HarmonicNumber[-1 - n1 + n2]*Log[y] + 
     4*HarmonicNumber[-1 + 2*n2]*Log[y] - 
     4*HarmonicNumber[-1 - 2*n1 + 2*n2]*Log[y] + Log[y]^2))/
   ((-1)^(3*n2)*t^2*n1!*n2!*(-1 - 2*n1 + 2*n2)!), 
  n1 >= 0 && 2*n2 >= 1 && 1 + n1 <= n2, {n1, n2}]}
\end{verbatim}
together with the message how the contour was closed : 
\begin{verbatim}
z1->R ( Re z1 > -1/2 )
z2->R ( Re z2 > -1/4 )
\end{verbatim}
Here the original choice how to close the contours was changed.
 We can now check the result numerically: $$\verb|DoAllMBSums[dim2sum, 50, Lk] // N|$$ gives
\begin{verbatim}
-0.0917188 eps
\end{verbatim}
in agreement with
\begin{verbatim}
{-0.0917189 eps, {8.79914*10^-6 eps, 0}}
\end{verbatim}
from \verb|MBintegrate[{dim2int}, Lk]//N|.

Further, we give examples of two potential errors. The
\begin{verbatim}
dim2sum = MBIntToSum[dim2int, Lk, {z2 -> L, z1 -> L}]
\end{verbatim}
prints
\begin{verbatim}
z2->R ( Re z2 > -1/4 )
Unable to found correct contour for z1.
\end{verbatim}
and returns \verb|{}|, while
\begin{verbatim}
dim2sum = MBIntToSum[dim2int, {t -> -10}, {z1 -> L, z2 -> L}]
\end{verbatim}
prints
\begin{verbatim}
Found c = -x (not a number): please complete kinematic's list.
Unable to find correct contour for z1.
\end{verbatim}
and returns \verb|{}|.

\subsection{Massive on-shell planar double box}

The next example is the massive on-shell planar double box diagram; see the example file \verb|2lbox.nb| at \cite{gluza-ustron-break:2015}.
The integral was studied in \cite{Smirnov:2001cm}.
Our kinematics is  $p_i^2=m^2$ for $i=1,2,3,4$, $p_1+p_2=-p_3-p_4$):
\begin{eqnarray}\label{I2l}
I & = &  e^{2\gamma_E \epsilon}\int \frac{d^{d}k_{1}}{i\pi^{d/2}}\frac{d^{d}k_{2}}{i\pi^{d/2}}  
  \frac{1}{	(k_{1}^2-m^2) 
  			(k_{1}+p_{1})^2 
  			[(k_{1}+p_{1}+p_{2})^2-m^2]
  			}\nonumber\\
  & &		\times	\frac{1}{(k_{1}-k_{2})^2(k_{2}^2-m^2)
  			[(k_{2}+p_{1}+p_{2})^2-m^2]
  			(k_{1}-p_{3})^2
  			}.
\end{eqnarray}
The MB-representation of (\ref{I2l}) agrees with \cite{Smirnov:2001cm}. 
Further, we will use the MB-representation (\ref{I2l}) in \verb|example7.nb| from \cite{Katowice-CAS:2015} in variable \verb|fin| 
without the 
$e^{2\gamma_E \epsilon}$ factor, which will be added later. Next the analytical continuation of \verb|fin| in $\epsilon\to 0$ as well as 
the expansion 
of the resulting integrals around $\epsilon=0$ up to $\epsilon^{-2}$ is obtained with \verb|MB| 
v.1.2.:
\begin{verbatim}
int1 = MBint[((-x)^(z1 + z6)*Gamma[-z1]^3*Gamma[1 + z1]*
   Gamma[-z6]^3*Gamma[1 + z6])/(2*eps^2*s^2*t*Gamma[-2*z1]*
   Gamma[-2*z6]), {{eps -> 0}, {z1 -> -1/2, z6 -> -1/2}}]
\end{verbatim}
with $x = m^2/s$, $(p_1+p_2)^2=s$, $(p_1+p_3)^2=t$.
We now produce sums from the MB-integral above, that are convergent at 
\begin{equation} \label{Lk2}
\verb|Lk = {s -> -1/5, x -> -5, t -> -1/10}|.
\end{equation}
Then $$\verb|s1 = MBIntToSum[int1, Lk, {z1 -> R, z6 -> R}]| $$ gives
\begin{verbatim}
{MBsum[((-1)^(-n1 - n2)*(-x)^(-n1 - n2)*n1!^2*n2!^2)/
   (2*eps^2*s^2*t*x^2*(1 + 2*n1)!*(1 + 2*n2)!), 
  n1 >= 0 && n2 >= 0, {n1, n2}]}
\end{verbatim}
together with the message how the contour was closed (the original choice is unchanged): 
\begin{verbatim}
z1->L ( Re z1 < -1/2 )
z6->L ( Re z6 < -1/2 )
\end{verbatim}

%The user can check that the same result can be obtained with \verb|dim1sum = MBIntToSum[dim1int, {}, {z1 -> L}]|.
The numerical check: $$\verb|DoAllMBSums[s1, 5, Lk] // N|$$ gives
\begin{verbatim}
-4.68459/eps^2
\end{verbatim}
in agreement with
\begin{verbatim}
{-4.6846/eps^2, {0.000453913/eps^2, 0}}
\end{verbatim}
from \verb|MBintegrate[{int1}, Lk]| and with the analytical result in \cite{Smirnov:2001cm}.

The two-dimensional sum in \verb|s1| can be written as a squared one-dimensional and  can be summed-up with Mathematica 9.0:
\begin{verbatim}
(8 ArcSin[1/(2 Sqrt[x])]^2)/(eps^2 s^2 t (-1 + 4 x))
\end{verbatim}
in numerical agreement with the above.

\subsection{Massless on-shell planar double box}

Next we consider the on-shell massless planar double box 
(\ref{I2l}) with $m=0$; see \cite{Smirnov:1999gc}.
Our kinematics is $p_i^2=0$ for $i=1,2,3,4$ and $p_1+p_2=-(p_3+p_4)$:
\begin{eqnarray}\label{I2lm0}
I & = &  e^{2\gamma_E \epsilon}\int \frac{d^{d}k_{1}}{i\pi^{d/2}}\frac{d^{d}k_{2}}{i\pi^{d/2}}  
  \frac{1}{	k_{1}^2 
  			(k_{1}+p_{1})^2 
  			(k_{1}+p_{1}+p_{2})^2
  			}\nonumber\\
  & &		\times	\frac{1}{(k_{1}-k_{2})^2 k_{2}^2 
  			(k_{2}+p_{1}+p_{2})^2
  			(k_{1}-p_{3})^2
  			}\nonumber\\
  & = &		\frac{K(x,\epsilon)}{(-s)^{2+2\epsilon}(-t)},	
\end{eqnarray}
where $x = t/s$, $(p_1+p_2)^2=s$, $(p_1+p_3)^2=t$; see the example file {\tt 2lbox-m0.nb} at 
\cite{gluza-ustron-break:2015}. The MB-representation for (\ref{I2lm0}) is obtained using \verb|AMBRE| v.1.2, based 
on 
the derivation of the MB-representation in \verb|example7.nb| (available at \cite{Katowice-CAS:2015}). From the four-dimensional 
MB-representation stored in variable \verb|fin| we removed the $1/(-s)^{2+2\epsilon}(-t)$ factor and obtained $K(x,\epsilon)$, 
stored in variable \verb|Kfin|.
Compare this with the five-dimensional MB-representation for $K(x,\epsilon)$ in \cite{Smirnov:1999gc}.
Next, the analytical continuation of \verb|Kfin| at  $\epsilon\to 0$ and the expansion of the resulting integrals around $\epsilon=0$ up to 
$\epsilon^0$ are obtained with \verb|MB| v.1.2. 
Finally, with \verb|barnesroutines| v.1.1.1 \cite{mbtools} one may reduce the dimensionality of 
the MB-integrals. 
One obtains several one-dimensional MB-integrals. 
Next, these MB-integrals are transformed into one-dimensional sums. 
We checked both 
the MB-integrals and sums to be in good numerical agreement at $x = 1/15$ with the analytic result for $K(x,\epsilon)$ in 
\cite{Smirnov:1999gc}. 
There were no sums (because zero-dimensional MB-integrals) to be done for the $\epsilon^{-2}$-term, and we obtained the analytic result 
already. 
The sums for the $\epsilon^{-1}$-term and the $\epsilon^0$-term could be done with advanced tools by J. Bl\"umlein and C. Schneider. They 
agree analytically with $K(x,\epsilon)$ in \cite{Smirnov:1999gc}.

%############################################################################
\section*{Acknowledgements}
We would like to thank J. Bl\"umlein, I. Dubovyk, J. Gluza and C.~Schneider for numerous helpful discussions.
% This program summary was also presented by J. Gluza at the conference ``Matter to the 
% Deepest'', Ustron, September 2015, and a shortened text version was prepared for the proceedings, to appear in APP.
The research project is partly supported by the Polish 
National Science Centre
(NCN) under the Grant Agreement No. DEC-2013/11/B/ST2/0402.

%\clearpage
% 
% \bibliographystyle{elsarticle-num} % with titles, preferred
% % not used here: \bibliographystyle{utphys_spires} % no titles
% \bibliography{2loops} % 2loops}

\end{document}